# Statistical Mechanics of Integral Membrane Protein Assembly

Karim Wahba, David J. Schwab, and Robijn Bruinsma
*Department of Physics, University of California, Los Angeles, CA 90095, USA*

**During the synthesis of integral membrane proteins (IMPs), the hydrophobic amino acids of the polypeptide sequence are partitioned mostly into the membrane interior and hydrophilic amino acids mostly into the aqueous exterior. We analyze the minimum free energy state of polypeptide sequences partitioned into α-helical transmembrane (TM) segments and the role of thermal fluctuations using a many-body statistical mechanics model. Results suggest that IMP TM segment partitioning shares important features with general theories of protein folding. For random polypeptide sequences, the minimum free energy state at room temperature is characterized by fluctuations in the number of TM segments with very long relaxation times. Simple assembly scenarios do not produce a unique number of TM segments and jamming phenomena interfere with segment placement. For sequences corresponding to IMPs, the minimum free energy structure with the wildtype number of segments is free of number fluctuations due to an anomalous gap in the energy spectrum, and simple assembly scenarios produce this structure. There is a threshold number of random point mutations beyond which the size of this gap is reduced so that the wildtype groundstate is destabilized and number fluctuations reappear.**

    Anfinsen established in a landmark study that the three-dimensional structure of globular proteins is determined by their primary amino acid sequences and that this structure is a minimum free energy state [1]. Integral membrane proteins (IMPs) such as ion channels, ion pumps, porins, and receptor proteins, do not easily lend themselves to Anfinsen's method, and whether or not assembled IMPs represent global free energy minima is not known [2]. The focus of this paper is on one of the most common IMP structures: bundles of, typically, seven to twelve transmembrane (TM) α–helices (Figure 1, inset). The helices consist of around 20-25 mostly a-polar amino acid residues linked outside the membrane by short, disordered polypeptide sequences of mostly hydrophilic amino acids. The TM segments can exist as stable entities inside the membrane in the absence of the bundle structure as the characteristic energy scale of tertiary structure formation is significantly lower than the formation free energy of the α–helices [3]. The identification of prospective α–helical TM

segments of a polypeptide sequence is an easier task than the prediction of the secondary structure of a globular protein. One procedure starts from a *hydropathy plot*, a plot of the free energy gained by transferring a certain number of successive amino acids of the primary sequence from aqueous environment into the membrane interior in the form of an α-helix, as a function of the start site of the segment. TM segment insertion free energies are assigned on the basis of an empirical hydrophobicity scale for the different amino acids [4]. Locations along the plot where the free energy gain for segment formation exceeds a certain threshold are possible start sites for TM segments. The hydrophobicity $\delta$ of individual amino acids in earlier hydropathy plots was obtained from solubility studies of amino acids in organic solvents, with considerable variation between different scales. In a commonly used scale [5], the variation of $\delta$ values was about 15 kcal/mole and the hydropathy plot values varied roughly between –40 and +30 kcal/mole. Segment placement for IMP sequences based on hydropathy plots is relatively straightforward and reproduces reasonably well the locations of α–helical segments of IMPs as obtained from x-ray structural studies [6]. More elaborate hidden Markov models, trained on known IMP structures, produce quite accurate structures [7, and references therein].

In the minimum free energy state, *fluctuations* in the number of TM segments – which would interfere with IMP functionality – can be neglected if the thermal energy $k_BT$ is small compared to the free energy difference $\delta E$ between structures with different numbers of segments. According to a crude statistical argument[*], the typical $\delta E$ of a long, generic (i.e., randomly picked) polypeptide sequence is of the order of $\langle \delta^2 \rangle^{1/2} L^{3/2} / N$, with $\langle \delta^2 \rangle^{1/2}$ the RMS variation of the hydrophobicity scale for the residues of the sequence, $L$ the mean TM segment length and $N$ the chain length. In the limit of large $N$, segment number fluctuations are thus unavoidable but for $N \sim 200$, a $\langle \delta^2 \rangle^{1/2}$ of the order of 8 kcal/mole, and $L \sim 20$, $\delta E$ is of the order of 3.6 kcal/mole and the effect of segment number fluctuations would be minor at room temperature ($k_BT \sim 0.59$ kcal/mole).

---

[*] Assume that the hydrophobicities of residues $j=1,…,N$ adopt the values $\delta S(j)$ where $S(j)= \pm 1$ with equal probability 1/2. Assume the mean hydrophobicity is absorbed in $\mu$. A TM segment of size $L$ starting at site $k$ has insertion energy $\Delta G(k) \sim \delta M(k) - L\mu$ with $M(k) = \sum_{j=k}^{k+L-1} S(j)$ the sum of $L$ random variables. $M$ is an approximately Gaussian random variable of zero mean and variance $\langle M^2 \rangle$ equal to $L$. A random chain of length $N$ corresponds to roughly $N/L$ independent tries of this random variable. The average spacing $\delta E$ between entries in the distribution of outcomes is then $\delta \langle M^2 \rangle^{1/2} / (N/L) \sim \delta L^{3/2} / N$.

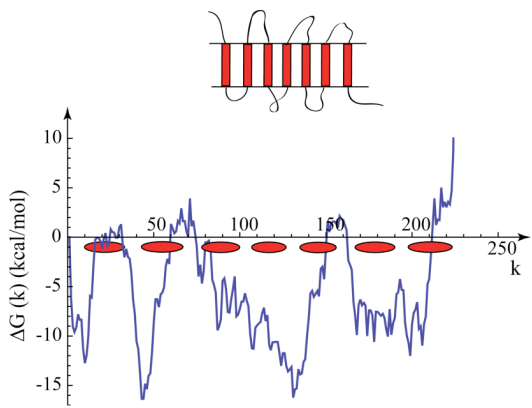

Figure 1: Insertion free energy $\Delta G(k)$ of a transmembrane $\alpha$-helical segment (Eq. (1) with $\mu = 0.7$ kcal/mol and $L_\alpha = 26$) for different values of the location $k$ of the first amino acid. The insertion free energy was computed using the hydrophobicity scale of Ref. [9] for the membrane protein bacteriorhodopsin (bR). The ellipses indicate the groundstate of the 7-segment wildtype structure. The inset schematically shows seven ordered $\alpha$-helical segments connected by disordered linker sections.

Synthesis of IMPs by ribosomes takes place on the surface of the endoplasmic reticulum where active clusters of proteins – translocons – thread unfolded, nascent polypeptide sequences through a transmembrane channel [8]. The translocon sequentially recognizes hydrophobic sections along the primary sequence and partitions them into the membrane. A remarkable study by Hessa et al. [9] showed that the translocon partitioning probabilities of different amino acid repeat sequences appear to follow equilibrium Boltzmann statistics. From measured probabilities, they established a new hydrophobicity scale that is appropriate for translocon partitioning. This scale has a relatively small range of hydrophobicity values, roughly from –0.6 to +3.5 kcal/mole, depending also on the location of the amino acid within the segment [10], while insertion free energies of IMP TM segments are in the range of –5 to +4 kcal/mole. If the insertion energies of 7–12 segments are uniformly distributed over this range, then $\delta E$ would *not* be large compared to the thermal energy. Similarly, if for a generic sequence one takes $\langle \delta^2 \rangle^{1/2}$ to be of the order of 1.0 kcal/mole in the earlier estimate, then $\delta E$ is of the order of the thermal energy. Both arguments suggest that segment thermal number fluctuations cannot be neglected in the minimum free energy state, notwithstanding the fact that IMP functionality requires a well-defined number of TM segments.

This paper applies methods of statistical mechanics to examine segment number fluctuations in the minimum free energy state of polypeptide sequences inserted into a membrane, as well as the conditions under which sequentially assembled structures can be expected to lead to a minimum free energy state with a well-defined number of segments. Under certain assumptions, the problem of placing variable sized, non-overlapping TM segments at finite temperature along a polypeptide sequence, with an insertion energy obtained from some hydrophobicity scale, can be reduced to a many-body problem of one-dimensional statistical mechanics in the grand-canonical ensemble whose minimum free energy state can be obtained by the transfer matrix method. When this model is applied to generic polypeptide sequences in combination with the Hessa scale, one finds that the minimum free energy state at room temperature indeed is characterized by strong thermal segment number fluctuations. The free energy barrier that has to be overcome to actually change the number of TM segments is, unlike $\delta E$, large compared to the thermal energy, even on the Hessa scale. That means that a minimum free energy state with strong thermal number fluctuations would, over a large range of time scales, actually correspond to a *glassy* state with a structure determined by assembly history. For polypeptide sequences corresponding to actual IMPs, there is however a "gap" in the excitation energy spectrum of the groundstate if the number of inserted segments $P$ equals the wildtype number of segments $P_w$. Because of this gap, IMPs are, in a state of minimum free energy, practically free of number fluctuations at room temperature. When $P$ is *not* equal to $P_w$, number fluctuations once again play an important role in the minimum free energy state.

We investigated the accessibility of the minimum free energy state by simple *sequential* assembly scenarios, for example by a translocon-type device. For the specific polypeptide sequences associated with IMPs, sequential assembly reproduces the number of segments of the groundstate as long as the number of segments $P$ is equal to or less than $P_w$, while for $P$ larger than $P_w$, jamming-type phenomena cause sequential assembly to produce non-unique segment placements. Only the structures produced by sequential assembly with $P = P_w$ reduce, at room temperature, to a minimum free energy state. In the presence of point mutations, the stabilizing anomalous energy gap shrinks as the number of random point mutations increases until a threshold is reached marked by growth of number fluctuations.

The contrast between the wildtype and random sequences in terms of thermodynamics and assembly kinetics is rather similar to that between the glassy "molten globule" state of collapsed generic polypeptide sequences in bulk solutions and the "designed" folded state of globular proteins at the lowest point of the folding "funnel" [11]. This folded state is usually free of large-scale, destabilizing thermal fluctuations, and is accessible from the unfolded state by rapid assembly kinetics. This suggests that IMPs and globular proteins can be described by a common general phenomenology.

## Methods

**The Model.** Assume a polypeptide sequence composed of $N$ hydrophilic and hydrophobic residues. We need to determine the statistical likelihood for an arbitrary sequence of TM segments of variable length and location placed in a hydrophobic environment, connected by disordered linker segments of residues placed in aqueous environment. Let the start site of a particular TM segment be denoted by the integer index $k$ and the number of TM residues by $L_\alpha$ with $\alpha$ indexing the set of observed different TM sizes. The model assigns a segment insertion free energy

$$\Delta G_\alpha(k) = \sum_{j=k}^{k+L_\alpha-1} (\delta(j) - \mu) \quad (1)$$

with $\delta(j)$ the hydrophobicity of residue $j$, for which we use the scale of Reference [9]. Thermodynamic changes of the environment that shift the zero of the hydrophobicity scale are included by the parameter $\mu$. The weakly attractive tertiary interaction free energy of one segment with the other segments of an IMP bundle can be absorbed as a contribution to $\mu$. Physically, $\Delta G_\alpha(k)$ can be viewed as the external potential energy of a TM segment of length $L_\alpha$ sliding along the primary sequence. Figure 1 shows $\Delta G_\alpha(k)$ with $\mu = 0.7$ kcal/mole and $L_\alpha = 26$ residues for the case of the well-studied integral membrane protein bacteriorhodopsin (bR), a 7-TM segment protein found in the outer membrane of *Halobacterium salinarium*. The model assumes "excluded-volume" repulsion between TM segments, i.e., segments are not allowed to overlap but the end site of one TM segment can be adjacent within two residues to the start site of the next TM segment with no free energy penalty. Specifically, for a rod of species $\alpha$ starting at site $j$ followed by another rod (of any species) starting at site $k > j$, the interaction potential is

$$V_\alpha(k-j) = \begin{cases} 0 & k-j \geq L_\alpha + 2 \\ \infty & k-j < L_\alpha + 2. \end{cases} \quad (2)$$

It should be noted that the aim of this model is to study the large-scale statistical mechanical properties of TM segment structures. Effects such as variations in the effective hydrophobicity of a residue due to correlations with neighboring residues, or linker-mediated, longer-range interactions between adjacent segments, which are not part of the model, may well affect the details of TM placement. However, they are not expected to be essential for the large-scale statistical mechanical properties that we are investigating in this study.

**Recursion Relations.** The Boltzmann statistical weight for the formation of a single TM segment of species $\alpha$ starting at site $k$ is defined as $e^{-\beta \Delta G_\alpha(k)}$ with $\beta = 1/k_B T$. The Boltzmann statistical weight $\rho_\alpha(k)$ for the site $k$ to be the start of a TM segment of length $L_\alpha$ as part of an *ensemble* of other segments is expressed as

$$\rho_\alpha(k) = e^{-\beta \Delta G_\alpha(k)} \Xi_\alpha^F(k) \Xi_\alpha^B(k) / \Xi \quad (3)$$

The term $\Xi_\alpha^F(k)$ represents the "forward" Boltzmann statistical weight of all possible TM segment distributions located anywhere between sites 1 and $k$ given that that there is a TM segment of size $L_\alpha$ that starts at site $k$. Similarly $\Xi_\alpha^B(k)$ represents the "backward" weight, while $\Xi$ is the overall normalization. Once $\rho_\alpha(k)$ has been determined, the mean number of TM segments $\rho_{TM} = \sum_\alpha \sum_k \rho_\alpha(k)$ can be obtained as a function of $\mu$. The slope $\chi = d\rho_{TM}(\mu)/d\mu$ at the values of $\mu$ where $\rho_{TM}(\mu)$ is equal to an integer $P$ plays the role of the *susceptibility* of a $P$ segment structure to thermal number fluctuations[†]. For a segment of length $L$, thermal number fluctuations become important when $\chi$ is of the order of $L/k_BT$ or larger. Of interest to compute also is the residue occupancy $\sigma(k) = \sum_\alpha \sum_{j=k-L_\alpha+1}^{k} \rho_\alpha(j)$ defined as the probability that a residue $k$ is part of a TM segment of any allowed size. A plot of $\sigma(k)$ shows the most probable locations of the TM segments.

Mathematically, the problem of computing TM placement probabilities has now been reduced to the computation of the grand canonical partition function $\Xi$ and the site-specific[1] "one-sided" partition functions $\Xi_\alpha^F(k)$ and $\Xi_\alpha^B(k)$ of a one-dimensional, multi-species liquid of variable-sized hard rods subject to an external potential. In the transfer matrix method, one first breaks up $\Xi_\alpha^F(k)$ as a sum over the different possible values of the distance $k - j$ (in residues) between a segment of size $L_\alpha$ starting at $k$ and a neighboring segment starting at site $j$ with $1 \le j < k$:

$$\Xi_\alpha^F(k) = e^{\beta \Delta G_\alpha(k)} \sum_\gamma \sum_{j=1}^{k-1} \Xi_\gamma^F(j) W_{\alpha,\gamma}(k-j) \quad (4)$$

The term $W_{\alpha,\gamma}(k-j) = \exp(-\beta V_\gamma(k-j))$ takes into account the excluded volume interaction between two neighboring TM segments of length $L_\alpha$ and $L_\gamma$ starting at sites $k$ and $j$ respectively. If the linker length obeys $k - j - L_\gamma < 2$, then $W = 0$ while $W = 1$ otherwise. Notice that in Eq. (4) one takes an annealed average over allowed TM segment sizes. This relation can be expressed as transfer matrix relation. Starting from the "initial condition" $\Xi_\alpha^F(1) = 1$, the values of $\Xi_\alpha^F(k)$ for $k > 1$ can be computed by forward iteration. A similar relation holds for the backward weights:

$$\Xi_\alpha^B(k) = e^{\beta \Delta G_\alpha(k)} \sum_\gamma \sum_{j=k+1}^{N} \Xi_\gamma^B(j) W_{\alpha,\gamma}(j-k) \quad (5)$$

which is reconstructed starting from $\Xi_\alpha^B(N) = 1$. Using these recursion relations it is possible to numerically reconstruct $\rho(\mu)$ under conditions of thermodynamic equilibrium for any given amino acid sequence. The transfer matrix method is closely related to hidden Markov models [7] while for the case that all segments have the same size, it reduces to the analytically soluble Percus model [12] of hard rods in an external potential.

---

[†] In the grand canonical ensemble, it corresponds to the second derivative of the thermodynamic potential with respect to the chemical potential.

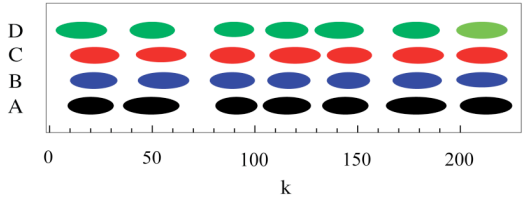

Figure 2: Comparison of the results of different segment placement methods for bR. Row A: measured segment locations from Ref. [13]; Row B: groundstate computed from the model for variable segment lengths at $\mu = 0.7$ kcal/mol and $L_\alpha = 26$; Row C: placement by sequential adsorption; Row D: linear sequential placement.

**Results**

**Ground State Stability and Thermal Fluctuations.** Figure 1 shows the bR groundstate structure, computed for the case that thermal fluctuations were "turned off" (i.e. the limit of large β), $\mu = 0.7$ kcal/mol and $L_\alpha = 26$, superimposed on the corresponding hydropathy plot with $L_\alpha = 26$. Segment start sites correspond to the local minima of the plot. Figure 2 compares this structure (row B) with the one reported by structural studies [13] shown as row A. The $\mu$ parameter was chosen to produce the best fit. The computed number, size and locations of the TM segments are in reasonable agreement with the reported structure. Figure 3 shows the mean segment number $\rho_{TM}(\mu)$ as a function of $\mu$ for three different temperatures. For very weak thermal fluctuations (Figure 3A), $\rho_{TM}$ has a discontinuous, staircase-like shape with steps at the integer values. A vertical step of the staircase represents the insertion of another TM segment, say to a state with $P$ segments. The subsequent horizontal width $\Delta\mu(P)$ measures the free energy change per amino acid required to add yet one more segment to the $P$-segment state and hence measures the thermodynamic stability of the $P$-segment state against changes in the number of segments. The $\Delta\mu(P)$ values for $P$ equal to two, three, four, and five are less than 0.1 kcal/mole. When $k_BT$ is increased to 0.2 kcal/mole (about $100^0 K$, Figure 3B) these steps are nearly completely washed out, and when $k_BT$ is increased to room temperature (Figure 3C), steps with $P$ equal to one and six are smeared out as well. Note that $\rho_{TM}(\mu)$ now is a smoothly continuous function with a typical susceptibility $\chi$ – given by the slope – in the range of $L/k_BT$. The exception is the seventh, step which has "survived" as a section with a slope that is practically zero at the center. Thermal number fluctuations can be neglected only in this $\mu$ interval. The $P = 7$ structure happens to correspond to the wildtype structure of bR. We repeated the calculation for random (i.e., randomly shuffled) bR sequences, with typical results shown in Figures 4A and 4B. At room temperature, the susceptibility $\chi$ is now of the order of $L/k_BT$ in the same range of $\mu$ values where the bR sequence had a plateau. Figures 5A and 5B compare the occupancies $\sigma(k)$ of the bR and random bR sequences at room temperature for $\mu = 0.7$ kcal/mole. For the bR sequence, the locations of the seven segments in the minimum free energy state remain quite well defined, with occupancies mostly close to one or zero. The smearing of the occupancy at the edges of the sixth peak is due to fluctuations in location and size of the sixth segment. The occupancy pattern for a random bR sequence shows an ill-defined placement pattern with occupation probabilities adopting a range of values.

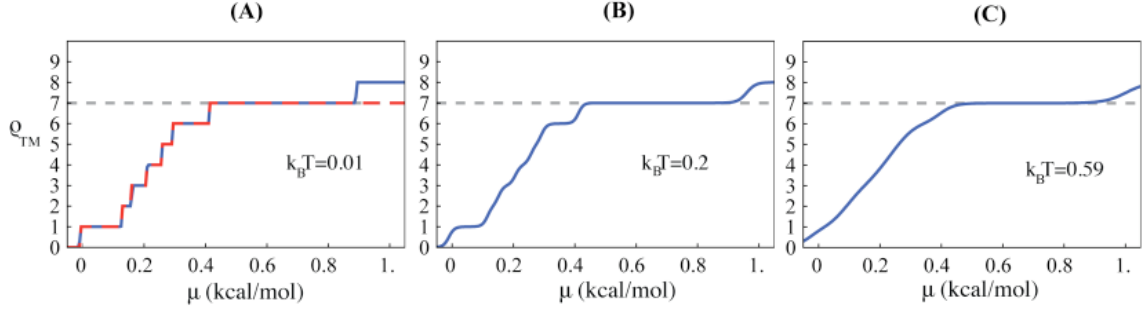

Figure 3: Mean number $\rho_{TM}$ of TM segments of bR as a function of the average insertion free energy gain $\mu$ per amino acid for different temperatures. (A) $k_BT = 0.01$ kcal/mole. The (red) dashed curve shows the mean number $\rho_{SA}$ of TM segments placed by sequential adsorption. For $\mu$ less than around 0.85 kcal/mole the two plots coincide but for $\mu$ larger than 0.85 kcal/mole $\rho_{SA}$ no longer increases. (B) $k_BT = 0.2$ kcal/mole. Only the 1-TM, 6-TM, and 7-TM segment structures have zero slopes at the respective center of the sections. (C) Room temperature ($k_BT = 0.59$ kcal/mole). Only the 7-segment structure has a zero slope.

In order to interpret occupancy profiles of this form, it is useful to overlay them on the hydropathy plot, as is done in Figure 6 for the same random bR sequence used in Figure 5. The thermal energy $k_BT$ was set to 0.1 kcal/mole and $\mu$ to 0.57 kcal/mole. This occupancy pattern is the superposition of occupancy patterns corresponding to four, respectively, five segments. In the 5-segment state (bottom of Figure 6), the last two segments occupy the two minima of the hydropathy plot indicated by circles. In the 4-segment state (top of Figure 6) one TM segment is placed with starting site either on the first circle or on the square, two nearly degenerate minima of the hydropathy plot. The energy differences between these three states are comparable to 0.1 kcal/mole so that all three states contribute at that temperature to the statistical ensemble and the occupancy plot is the superposition of the three states. There are thus both segment number fluctuations as well as large-scale positional fluctuations in this minimum free energy state.

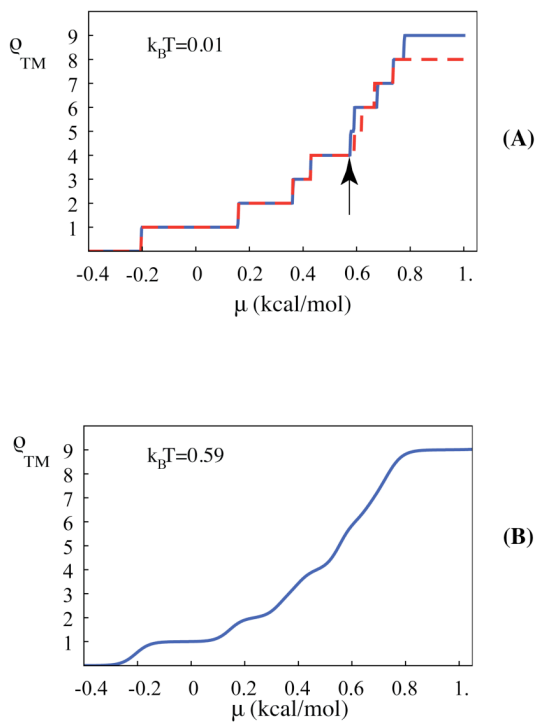

Figure 4: Mean number of TM segments $\rho_{TM}$ for a randomly shuffled bR sequence as a function of $\mu$. (A) $k_BT = 0.01$ kcal/mole. The plot of $\rho_{SA}$ for sequential adsorption (dashed red line) starts to deviate from $\rho_{TM}$ at the arrow. (B) Room temperature. Structures with more than one segment suffer from strong number fluctuations (up to the 9-TM segment structure). The corresponding occupancy plot at the site of the arrow is shown in Figure 6.

To check whether these results were specific for bR, we repeated the analysis for five 7-TM segment proteins and five 12-TM segment proteins (Table 1, column 3). In all cases, $\Delta\mu(P_w)$ was anomalously large and only the wildtype groundstate configuration survived at room temperature. Figures 7A and 7B show the cases of diacylglycerol kinase and cytochrome C oxidase, 3-TM and 12-TM segment IMPs respectively. Note that for the 12-TM segment proteins, the size of the wildtype stability interval $\Delta\mu(P_w)$ is particularly pronounced compared to the other stability intervals.

**Assembly Robustness.** In order to transform the 5-segment state of Figure 6 into one of the two 4-segment states, the fifth segment must be pulled out of the membrane. The mean free energy barrier for pull-out can be estimated as $\mu L$ or about 15 kcal/mole for $\mu$ equal to 0.6 kcal/mole. An Arrhenius estimate of the rate of segment pull-out by thermal fluctuations indicates that this would require macroscopic time scales[‡]. Segment number fluctuations are in general so slow that states whose number susceptibility $\chi$ approaches $L/k_BT$ are unlikely to be in thermodynamic equilibrium on laboratory time scales.

---

[‡] For an Arrhenius rate with an attempt frequency $k_BT/\eta_m d^3$ with $d$ the membrane thickness of 50 Angstrom and $\eta_m$ a membrane viscosity of 0.1 in SI units, the time scale for removing the last segment by thermal fluctuations would be in the range of 10 seconds assuming a 15 kcal/mole activation barrier.

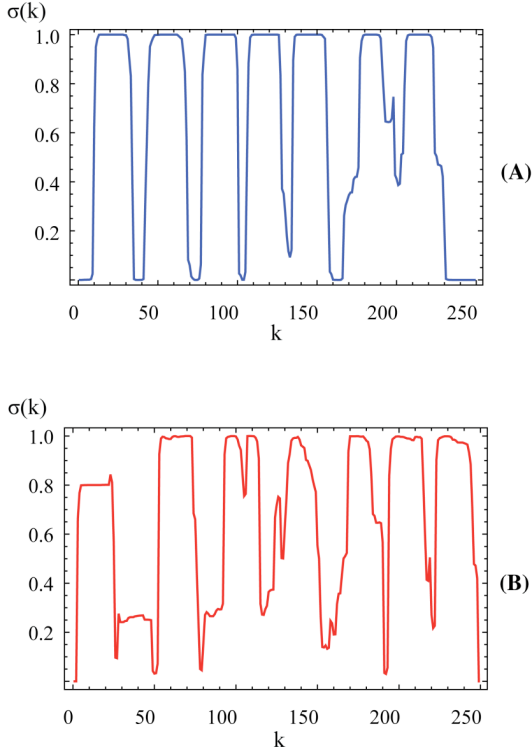

Figure 5: Plot of the probability σ(k) that amino acid k is part of a TM segment at room temperature for $\mu = 0.7$ kcal/mol. (A) Wildtype bR sequence. The sixth segment is subject to location and length fluctuations. There are no number fluctuations. (B) Shuffled bR sequence. In addition to strong location/size fluctuations there are also segment number fluctuations.

In this section we will examine TM segment structures that are not in full thermodynamic equilibrium by assuming the case that segment number fluctuations do not take place and that the number of TM segments is fixed during the initial assembly. Size and location fluctuations remain possible. We will inquire under which conditions the assembled state will be an approximate minimum free energy state for two simple sequential assembly scenarios. Row D in Figure 2 shows the result of the assembly of the bR sequence in a scenario where, starting at one end, one sweeps through the sequence placing a new TM segment on the first available low-energy binding site not covered by the previous segment, demanding only that the binding energy exceeds a certain threshold. By carefully tuning this threshold, this "linear sequential" placement of the TM segments can be made to agree both with the measured structure (row A) and the computed groundstate (row B). Row C shows the result of assembly by "sequential adsorption", where one places the first TM segments at the minimum of $\Delta G_\alpha(k)$ with respect to k and α, then searches for the next lowest value of $\Delta G_\alpha(k)$ that is not blocked by the first segment, and repeating this procedure as long as sites with negative $\Delta G_\alpha(k)$ can be located for the given μ. All four rows place the segments in approximately the same locations. Figure 3A shows the mean segment number $\rho_{SA}(\mu)$ obtained by sequential adsorption for the bR sequence as a dashed line. Sequential adsorption *exactly* reproduces $\rho_{TM}(\mu)$ up to and including $P = 7$ but then sequential adsorption halts while $\rho_{TM}(\mu)$ continues to increase. This "jamming" phenomenon is a familiar feature of studies of sequential adsorption in other systems [14]. The case of sequential adsorption of a random bR chain is shown in Figure 4A (dashed line). As expected from Figure 5, discrepancies between $\rho_{SA}(\mu)$ and $\rho_{TM}(\mu)$ appear at $P = 4$. We repeated this analysis for other proteins and always found that sequential adsorption reproduces the groundstate up to the wildtype number of TM segments, while random sequences encounter placement discrepancy for lower values of $\mu$ (Figures 7A and 7B compare $\rho_{SA}(\mu)$ (dashed line) and $\rho_{TM}(\mu)$ for cytochrome C oxidase and diacylglycerol kinase).

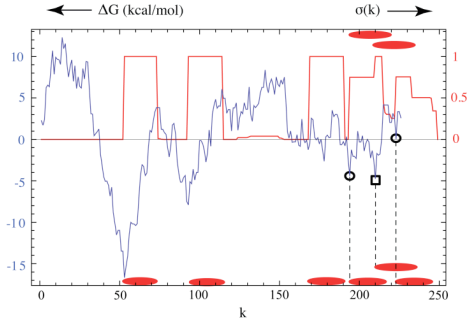

Figure 6: Occupancy plot for the randomly shuffled bR sequence at $\mu = 0.57$ kcal/mol overlaid on the hydropathy plot for $k_B T$ set at 0.1 kcal/mol. Bottom row of ellipses: structure of the 5-TM segment groundstate with the locations of the last two segment indicated by circles. The minimum at $k = 210$ (square) is blocked in this structure. Top: two alternative placements of the last segment in the two competing 4-TM segment states. The occupancy plot is the superposition of these three nearly degenerate states.

Recall that we found that the room temperature susceptibility $\chi$ for number fluctuations was negligible for $P = P_w$ at the center of the wildtype stability interval, so number fluctuations were not required for thermal equilibration. We conclude that simple assembly scenarios effectively *can* produce the unique minimum free energy state of IMPs with $P = P_w$. Structures with lower $\mu$ values, where sequential assembly also produced the correct groundstate, but now with $P$ less than $P_w$, *did* require segment number fluctuations for thermal equilibration. The earlier conclusion thus only holds for $P = P_w$. For shuffled IMP sequences with $\mu$ in the same range, sequential assembly scenarios are not consistent with each other and their final states cannot reach thermodynamic equilibrium without slow number fluctuations. This result suggests that random *mutations* could interfere with IMP assembly, which we will now investigate.

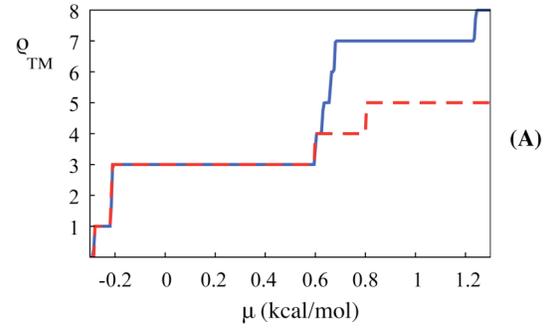

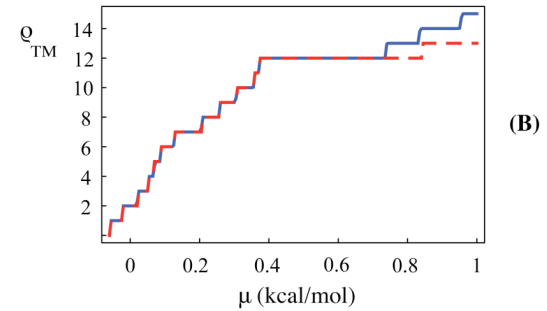

Figure 7: Mean number of TM segments $\rho_{TM}$ as a function of $\mu$. (A) Diacylglycerol kinase, a 3-TM segment protein. (B) Cytochrome C oxidase, a 12-TM segment protein. Dashed (red) lines indicate sequential adsorption density $\rho_{SA}$.

**Mutational Robustness.** The structure of many globular proteins is robust with respect to random point mutations [15]. In addition to the obvious advantage of preserving functionality in the presence of mutations, mutational robustness also increases the number of sequences that map to the same folding structure, thereby promoting diversification and "evolvability" [16]. Does the large energy gap that protects the groundstate of IMPs against "destructive" segment number fluctuations also provide robustness against mutations?

We computed the number of randomly chosen single point mutations (SPMs) required to produce a change in the groundstate number of TM segments, both for IMP sequences and their random analogs. The value of $\mu$ was fixed at the center of the stability gap $\Delta\mu(P_w)$ for the wildtype structure. We repeated this procedure a hundred times and computed the average number of SPMs (normalized by sequence length) to produce a change in the number of segments as well as the standard deviation. We then repeated this procedure for each protein with an ensemble of a hundred realizations of randomly shuffled sequences. The results are shown in Table 1, in columns 4 and 5. For the random sequences (column 5), one to five point mutations per hundred residues typically were sufficient to change the number of TM segments in the groundstate. For IMPs, the SPM threshold was systematically higher than that of the randomized sequences. For bR, and other 7-TM segment IMPs, the SPM threshold was about five times higher but for certain 12-TM segment IMPs, like lactose permease of *E. coli*, the SPM threshold enhancement was only a factor of two larger. Columns 4 and 5 show that there is some correlation between the thresholds of the wildtype and shuffled sequences and columns 3 and 4 show that there is some correlation between the thermodynamic stability interval $\Delta\mu(P_w)$ and the mutation threshold for most IMPs but also there are striking exceptions. An example is the bacterial protein glycerol-3 phosphate transporter (*E. coli*) that has the largest energy gap yet only modest mutational robustness. We conclude that the groundstate of wildtype IMPs are in most cases indeed significantly more stable against point mutations than the shuffled sequences but thermodynamic and mutational robustness are in general separate properties of an IMP.

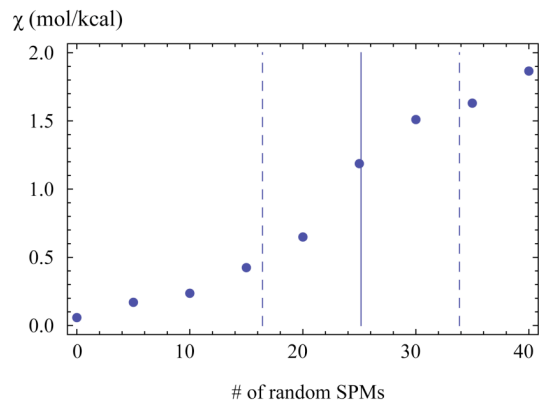

Figure 8: Susceptibility $\chi = d\rho_{TM}/d\mu$ for thermal segment number fluctuations of bR as a function of the number of randomly chosen single point mutations (SPMs) of the sequence. Each point is an average over 100 trials. The solid vertical line marks the threshold where the groundstate structure is destabilized by mutations, and corresponds to the mean given in Table 1. The error bars are shown as dashed vertical lines.

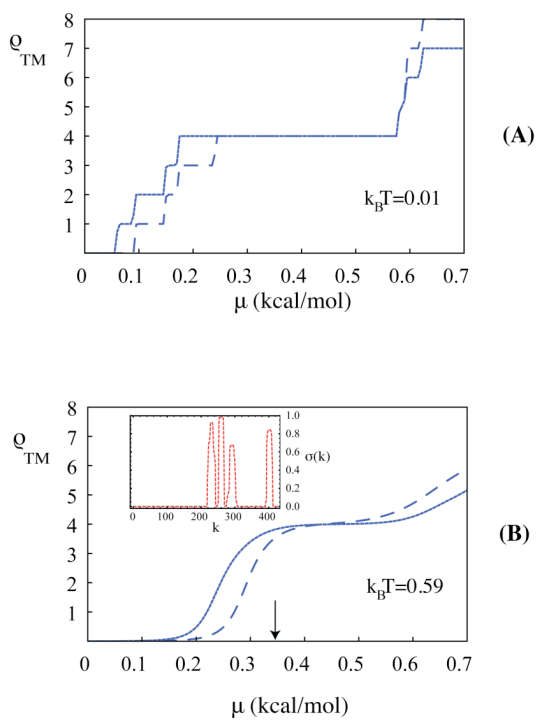

Figure 9: Effect of A322D mutation on the GABA$_A$ receptor α₁subunit. (A) $\rho_{TM}$ at $k_BT$ set at 0.01 kcal/mole. The mutant is indicated by the dashed line. (B) $\rho_{TM}$ at room temperature, with the mutant indicated by the dashed line. The inset shows the occupancy of the mutant at the value of $\mu$ = 0.34 kcal/mol indicated by the arrow. The mutation occurs in the third TM segment.

Is there a relation between the mutational threshold and susceptibility to segment number fluctuations? Figure 8 shows the average susceptibility for number fluctuations of bR at the center of the 7-segment interval as a function of the number of mutations. The mutation threshold of Table 1 is indicated as a vertical line with the dashed lines indicating the error bars. The threshold is the locus of a rapid rise of the susceptibility for number fluctuations. The mutation threshold thus marks both a change in the groundstate structure and an increased lack of stability against segment number fluctuations

**Conclusion**

According to the model presented in this paper, polypeptide sequences associated with actual IMPs *can* be assembled into minimum free energy structures by simple sequential assembly scenarios but this is not true for generic sequences. Assembly robustness is achieved by (i) an anomalously large gap in the energy excitation spectrum that prevents thermal number fluctuations, and (ii) by the absence of jamming-type phenomena for segment numbers equal to or less than the wildtype. Generic sequences of the same length and the same amino acid abundance as an IMP sequence are in a glassy state with a structure that depends on the details of the assembly history.

Is there experimental evidence for thermal number fluctuations in IMPs? If segment number fluctuations indeed do take place on laboratory time scales then this would show as a statistical uncertainty in the number of TM segment after IMP assembly. The TM helix formation of the GABA$_A$ receptor $\alpha 1$ subunit is destabilized by a particular point mutation, the A322D mutation, which causes a form of myoclinic epilepsy [17]. The wildtype GABA$_A$ receptor subunit is a 4-TM segment structure, and for the A322D mutant, the third segment fails to insert into the lipid bilayer about 33% of the time. Figure 9A shows $\rho_{TM}(\mu)$ computed for both the wildtype and the A322D mutant in the absence of thermal fluctuations. Note that the stability interval of the 4-TM segment structure of the mutant is noticeably shorter compared to the wildtype. Figure 9B shows $\rho_{TM}(\mu)$ of the wildtype and the A322D mutant at room temperature, with the occupancy plot inset. For $\mu$ near the value (indicated by the arrow in Figure 9B) where the mean number of segments is about 3.5, the susceptibility approaches $L/k_BT$. The A322D mutant should be characterized by strong segment number variations during assembly, as is indeed the case experimentally.

We close by noting that a model similar to the one discussed in this paper has been applied to the problem of the placement of nucleosomes along genomic DNA molecules. By comparing measured structures with the minimum free energy state computed for the model, it was established that the assembly of DNA-nucleosome fibers does generate a state of near minimum free energy [18], despite very large free energy barriers between structures with different numbers of nucleosomes. Because of the much greater length of the genome sequence, assembly frustration of the form shown in Figure 5, was unavoidable. The competing states appear to act as biological switches [19]. It would be interesting if artificial IMPs could be synthesized that – like the GABA$_A$ subunit – can exist in two alternative switch forms with different numbers of segments, for example by altering the amino acid sequence of an IMP, explicitly introducing assembly frustration of the form shown in Figure 5, and testing which of the competing structures is assembled by the translocon.

**Acknowledgments**
We would like to thank the NSF for support under DMR Grant No. 04-04507.

| Protein (PDB #) | # of TM segments $P$ | $\Delta\mu(P_w)$ (kcal/mol) | Mutation threshold (wildtype) | Mutation threshold (randomized sequence) |
|---|---|---|---|---|
| Bacteriorhodopsin (1BRD) | 7 | 0.47 | 0.101 ± 0.035 | 0.024 ± 0.012 |
| Sensory rhodopsin – *Anabaena* (1XIO) | 7 | 0.44 | 0.063 ± 0.022 | 0.014 ± 0.007 |
| Halorhodopsin (1E12) | 7 | 0.34 | 0.048 ± 0.025 | 0.017 ± 0.009 |
| Bovine rhodopsin (1F88) | 7 | 0.49 | 0.093 ± 0.032 | 0.025 ± 0.012 |
| Sensory rhodopsin II (1H68) | 7 | 0.63 | 0.103 ± 0.029 | 0.045 ± 0.021 |
| Bovine Cyto. C. Oxidase – III (1OCC) | 7 | 0.43 | 0.121 ± 0.043 | 0.036 ± 0.017 |
| Bovine Cyto. C. Oxidase – I (1OCC) | 12 | 0.36 | 0.069 ± 0.026 | 0.006 ± 0.003 |
| Glycerol -3 Phosphate Transporter – *E. coli* (1PW4) | 12 | 0.6 | 0.055 ± 0.028 | 0.007 ± 0.004 |
| (*P. denitrificans*) Cyto. C. Oxidase – I (1QLE) | 12 | 0.22 | 0.023 ± 0.011 | 0.004 ± 0.002 |
| Lactose Permease – *E. coli* (1PV6) | 12 | 0.19 | 0.020 ± 0.010 | 0.010 ± 0.005 |
| Multi drug resistance protein EmrD – *E. coli* (2GFP) | 12 | 0.28 | 0.021 ± 0.011 | 0.010 ± 0.005 |

**Table 1.** The second column gives the number of wildtype TM segments of eleven representative IMPs shown in column 1 with their corresponding PDB id. The third column gives the size of the stability interval $\Delta\mu$ for $P = P_w$ of the groundstate structure as computed from the model. The fourth column gives the average (over 100 runs) number of random single point mutations (SPMs) normalized by sequence length required to change the segment from its wildtype value for each protein, including standard error. Note that 1PW4 has the maximum thermodynamic stability $\Delta\mu$ but a low mutation threshold. The fifth column gives the average (over 100 runs) mutation threshold for each sequence after random shuffling (100 realizations). Note the correlation between columns four and five. For the last rows, the mutation threshold of the shuffled sequence is comparable to the wildtype sequence.